\documentclass[11pt]{article}

\pdfoutput=1
\usepackage[pdftex]{color}
\usepackage{amssymb}
\usepackage{amsthm}
\usepackage{amsmath}
\usepackage{latexsym}
\usepackage{amscd}
\usepackage{graphicx}
\usepackage[pdftex, colorlinks=true, citecolor=green]{hyperref}
\usepackage{lscape}
\usepackage{setspace}
\usepackage{multirow}

\newcommand{\ds}{\displaystyle}

\newcommand{\ben}{\begin{equation}}     %equation
\newcommand{\eeqn}{\end{equation}}
\newcommand{\bey}{\begin{eqnarray}}
\newcommand{\eey}{\end{eqnarray}}

\begin{document}

\begin{flushleft}
\noindent {\Large
\textbf{A mathematical model of reward and executive circuitry in obsessive compulsive disorder}
}
\\
\vspace{4mm}
\noindent  Anca R\v{a}dulescu$^{*,}\footnote{ Assistant Professor, Department of Mathematics, State University of New York at New Paltz; New York, USA; Phone: (845) 257-3532; Email: radulesa@newpaltz.edu}$, Rachel Marra$^{2}$
\\
\noindent $^1$ Department of Mathematics, SUNY New Paltz, NY 12561
\\
\noindent $^2$ Department of Astronomy, SUNY New Paltz, NY 12561
\end{flushleft}

\begin{abstract}
\noindent The neuronal circuit that controls obsessive and compulsive behaviors involves a complex network of brain regions (some with known involvement in reward processing). Among these are cortical regions, the striatum and the thalamus (which compose the CSTC pathway), limbic areas such as the amygdala and the hippocampus, and well as dopamine pathways. Abnormal dynamic behavior in this brain network is a hallmark feature of patients with increased anxiety and motor activity, like the ones affected by OCD. There is currently no clear understanding of precisely what mechanisms generates these behaviors.

We attempt to investigate a collection of connectivity hypotheses of OCD by means of a computational model of the brain circuitry that governs reward and motion execution. Mathematically, we use methods from ordinary differential equations and continuous time dynamical systems. We use classical analytical methods as well as computational approaches to study phenomena in the phase plane (e.g., behavior of the system's solutions when given certain initial conditions) and in the parameter space (e.g., sensitive dependence of initial conditions).

We find that different obsessive-compulsive subtypes may correspond to different abnormalities in the network connectivity profiles. We suggest that it is combinations of parameters (connectivity strengths between regions), rather the than the value of any one parameter taken independently, that provides the best basis for predicting behavior, and for understanding the heterogeneity of the illness.
\end{abstract}

\section{Introduction}

\subsection{The obsessive--compulsive disorder and its behavioral dynamics}

Obsessive-compulsive disorder (OCD) is a severe mental disturbance affecting around 2-3\% of the US population. Its symptoms are chronic intrusive thoughts (obsessions) and/or repetitive behaviors (compulsions), which can lead to significant impairment in psychosocial functioning~\cite{american2003apa, leonard1990childhood}. While medication plans and behavioral therapy may benefit some patients, 20-40\% of OCD patients -- often those most severely affected -- remain refractory to treatment~\cite{skoog1999,bourne2012mechanisms}.

According to the American Psychiatric Association, obsessions are repetitive, intrusive, and distressing thoughts, ideas, images, or urges that often are experienced as meaningless, inappropriate, and irrelevant, and persist despite efforts to suppress, resist, or ignore them~\cite{american2003apa}. Compulsions are repetitive, stereotyped behaviors and/or mental acts that are used to diminish the anxiety and distress associated with the obsessions~\cite{american2003apa}. It has been noticed that obsessions and compulsions are often linked by content. For example, obsessive recurrent thoughts about erring may trigger compulsive checking of the work, and obsessive contamination concerns may lead to compulsive washing routines~\cite{markarian2010multiple}. Although the DSM-IV criteria imply that it is possible for a person to experience compulsions without obsessions or vice versa, the vast majority of OCD patients have both obsessions and compulsions. Indeed, only 2.1\% of patients with OCD report predominant obsessions, 1.7\% report predominant compulsions, and more than 95\% reported both obsessions and compulsions on the Yale-Brown Obsessive Compulsive Symptom Checklist~\cite{foa1995dsm}.

Structural and functional neuroimaging converge on evidence that abnormalities of obsessive compulvie behavior arise from abnormal neural activity in a wide network of cortical and subcortical areas. This network consists primarily of the orbitofrontal cortex (OFC) and the anterior cingulate cortex (ACC) -- the striatum (the ventral striatum in particular, also known as the nucleus accumbens -- NAc), the thalamus, the amygdala (and other limbic regions), but also includes dopamine regulation from the ventral tegmental area (VTA) ~\cite{bourne2012mechanisms,menzies2008integrating,rotge2008provocation,rotge2010gray,haber2008cognitive}. \\

\noindent {\bf The orbitofrontal cortex (OFC)} is active in the regulation of emotional and fear responses, as well as in  positive and negative valence processing~\cite{kringelbach2004functional}. The OFC is also involved in behavioral planning and expected reward valuation. Dysfunctions of these processes are likely involved in the repetitive compulsions and driving obsessions of OCD, and may be represented by the altered OFC activity seen in this disease. Imaging (PET and fMRI) studies have shown increased activity in the OFC in OCD patients  both during resting state and with symptom provocation~\cite{bourne2012mechanisms}.\\

\noindent {\bf The anterior cingulate cortex (ACC)} plays a role in motivation and conflict monitoring, as well as determining discrepancies between desired and anticipated state~\cite{graybiel2000toward}. ACC has close connections to the motor cortex, and, together with the OFC, plays an important role in action selection subsequent to stimulus valuation. It has been suggested that altered activity in these two cortical areas contributes to dysfunctional signaling of conflict between the desired and the current state -- that may lead to the repetitive actions (compulsions) characteristic of OCD~\cite{bourne2012mechanisms,graybiel2000toward}.\\

\noindent {\bf The thalamus} is often viewed as a hub of information, being tightly connected with a wide network of brain areas. It has multiple functions, including that of processing sensory information and relaying it to cortical areas, as well as that of ``subcortical motor center''~\cite{evarts1969motor}. As part of the cortico-striato-thalamo-cortical circuit, the thalamus has been studied in OCD, with evidence showing increased activity in both resting state and with symptom provocation~\cite{mcguire1994functional}.\\

\noindent {\bf The striatum} (and more generally, the basal ganglia) has long been implicated (as part of the cortico-striato-thalamo-cortical loop) in the pathophysiology of obsessive-compulsive disorder and remains central to neurobiological models of OCD.  The sensorimotor, associative, and limbic loci in the basal ganglia are widely implicated in motor, cognitive, and emotional aspects of behavior, respectively~\cite{harrison2009altered}. The ventral striatum (nucleus accumbens) in particular plays an important role in processing rewarding and reinforcing stimuli. Obsessive-compulsive disorder patients may be less able to make beneficial choices because of altered nucleus accumbens activation when anticipating rewards. This finding supports the conceptualization of OCD as a disorder of reward processing and behavioral addiction.\\

\noindent {\bf Limbic regions} are critically involved in the association of external stimuli with emotional value and the acquisition of conditioned fear responses. In OCD, abnormal activity in limbic structures may underlie the stimulus-triggered  anxiety which often urges the patient to execute compulsions. While the amygdala and hippocampus have not received as much attention in the OCD literature as the areas of the CSTC loop, newer studies reveal their potentially critical contribution to the behavioral dynamics of the illness, and their potential implications to its treatment (e.g., through supporting hippocampal neurogenesis~\cite{bourne2012mechanisms}). Existing studies offer conflicting evidence on whether OCD patients present functional and structural limbic abnormalities when compared to healthy controls~\cite{kwon2003similarity,busatto2000voxel}. These discrepancies may arise from the heterogeneity in the pathophysiology of OCD, depending on severity and type of obsessions / compulsions~\cite{bourne2012mechanisms}. For example, a study using proton magnetic resonance spectroscopy~\cite{besiroglu2011involvement} reported amygdala and hippocampus abnormalities in patients with autogenous obsessions (which do not emerge in response to identifiable external stimuli), which were not found in patients with reactive obsessions (evoked by specific external stimuli).\\

\noindent {\bf The ventral tegmental area (VTA)} is known to play, via dopamine modulation, a critical role in motivation, reward-related behavior, attention, and multiple forms of memory~\cite{malenka2009molecular}. With its wide net of projections to a variety of fields, the activity of dopamine neurons confer motivational salience on the reward (via its connections with the nucleus accumbens), reevaluate it in light of new experiences (via the ties with the OFC), contribute to memory consolidation (via its ties with the hippocampus and amygdala), contributes to the reward-seeking motor control, to suppression of behaviors that compete with goal oriented actions and to control of attention (via its actions in the prefrontal cortex)~\cite{malenka2009molecular}.\\

\noindent It is not surprising that some of these areas are known to be critical to the processing of rewards and to movement execution. New studies have been investigating the relationship between abnormalities in reward processing and the symptoms of OCD. It has been suggested that patients with OCD may develop dependency upon compulsive behaviors because of the rewarding effects following reduction of obsession-induced anxiety ~\cite{figee2011dysfunctional}. We investigate some of these possibilities in our modeling work.  

The existing conflicts between results in the specialty literature suggest that OCD, like other mental illnesses, has a heterogeneous etiology, and that there may be multiple biophysical mechanisms underneath what is being addressed as a homogeneous disorder. Considering the heterogeneity of OCD may be most important when investigating regions only weakly implicated in OCD (such as the amygdala), but that may nonetheless significantly affect pathology~\cite{bourne2012mechanisms}.

It is in this context that mathematical modeling can be successfully used in exploring potential mechanisms for different OCD subtypes. In this study, we aim to verify some of the current specific hypotheses that relate OCD symptoms to disruptions of connectivity and functional dynamics within the reward/executive network. While most imaging studies present these effects as separate possible causes or triggers of obsessive-compulsive behavior, we rather suggest that it is a complex combination of all these factors (and possibly other, unknown ones) which govern the behavior of the network. Our mathematical model presents the advantage of allowing us to analyze theoretically the effects of some of these factors taken separately, but also in combination. These effects predict different network dynamic regimes, which can then be compared with the dynamic patterns seen in imaging time series under, and further, with behavioral dynamics observed in OCD patients.

The rest of the paper is organized as follows: In Section~\ref{projections}, we describe the known projections within our network of interest. In Section~\ref{hypotheses}, we list a few hypotheses that relate OCD symptoms to connectivity and functional problems in this network. In Section~\ref{mother_model}, we introduce our mathematical model and discuss how these hypotheses are reflected in the construction. In Section~\ref{results}, we analyze the effect on dynamics of changes to specific parameters of interest, considered separately and in combination. We focus on relating the position and type of the asymptotic attractors (equilibria and cycles, in this case) to long-term behavioral regimes found in OCD (e.g., oscillation between elevated amygdala due to anxiety, and elevated ACC when performing compulsive movement execution). Finally, in Section~\ref{discussion}, we discuss some of the broader implications of our model, as well as limitations and further work.

\subsection{The reward/executive network}
\label{projections}

Our circuit of interest is comprised of a few modules, whose coupled behavior governs the brain's emotional and cognitive responses to positive and negative rewards: the cortex, the limbic system, the basal ganglia, the thalamus and the dopamine system. Below, we describe the most important excitatory (glutamate) and inhibitory (GABA) pathways as well as the more subtle dopamine modulations delivered by the ventral tegmental area.

The main cortical areas implicated in this circuitry are the orbitofrontal cortex (OFC), and the anterior cingulate cortex (ACC). The \textbf{\emph{orbitofrontal cortex}} receives glutamate (excitatory) projections from the thalamus~\cite{bourne2012mechanisms}, as well as from limbic areas such as the amygdala (which projects primarily to the lateral division of the OFC) and the hippocampus (which connects promarily with the medial OFC)~\cite{sotres2004emotional,elliott2000dissociable,cavada2000anatomical}. The \textbf{\emph{anterior cingulate cortex}} also receives glutamate afferents from thalamus~\cite{gigg1992glutamatergic} and the medial OFC~\cite{elliott2000dissociable}, as well as from limbic areas (subiculum, a part of the hippocampus) through the mammilary tract.

Limbic areas are known to have important contributions to the reward circuitry. The amygdala and the hippocampus have been identified in particular to be responsible for emotional modulation of reward processing. The \textbf{\emph{amygdala}} is kept under inhibitory control both by the orbitofrontal cortex~\cite{rempel2007role} (provided primarily by lateral OFC ~\cite{elliott2000dissociable}), and by the ACC~\cite{etkin2011emotional,rempel2007role}. This inhibition is provided to central and basal regions of the amygdala via indirect excitation of inhibitory (GABA) interneurons in the ITC. The amygdala also receives glutamatergic input from other limbic areas (e.g., the hipppocampus~\cite{kishi2006topographical,richter2000amygdala}) and from the thalamus~\cite{bourne2012mechanisms,turner1991thalamoamygdaloid}. The \textbf{\emph{hippocampus}}, in turn, is connected via symmetric afferent and efferent pathways~\cite{swenson2006review} to the OFC (primarily to the medial parts~\cite{elliott2000dissociable}), to the amygdala and to the thalamus.

The basal ganglia are part of the cortico-thalamo-striato-cortical loop of the reward circuit, which appears to be most implicated in OCD abnormalities. The \text{\emph{striatum}} is the central piece of the CTSC circuit, with its D1 and D2 spine MRNs receiving inputs (often not symmetric) from a wide variety of sources. Both D1 and D2 MRNs receive  (via direct and indirect pathways, respectively) unbiased glutamate inputs from the thalamus and from the OFC~\cite{wall2013differential}, and also glutamate projections from the hippocampus~\cite{liddle2000immediate}. In addition, the limbic (i.e., anterior cingulate) cortex sends glutamate inputs primarily to D1~\cite{wall2013differential}, while the motor cortices send a stronger glutamate proportion of inputs to D2~\cite{wall2013differential}. The amygdala also sends glutamate projections~\cite{yager2015ins} to the ventral striatum (nucleus accumbens)~\cite{cho2013cortico}, including to the striatal shell ~\cite{fudge2002amygdaloid,fudge2004amygdaloid}, which expresses equally D1 and D2~\cite{yager2015ins}.

The \textbf{\emph{globus palidum internum}} (Gpi)~\cite{liddle2000immediate} receives direct inhibitory inputs from D2 striatal cells, as well as indirect excitatory inputs from the D1 striatal cells (via a pathway of two inhibitory and one excitatory connections, with relay stops in the globus palidus externum and in the subtalamic nucleus). The Gpi sends in turn glutamate inputs to the \textbf{\emph{thalamus}}~\cite{liddle2000immediate}, which also receives excitatory inputs from the OFC~\cite{bourne2012mechanisms}, from the ACC, through the mammilary bodies~\cite{bourne2012mechanisms,vogt1993neurobiology}, and from limbic regions: from the hippocampus, via mamillo-thalamic and hippocampothalamic pathways~\cite{vogt1993neurobiology}, and form the amygdala~\cite{bourne2012mechanisms,krettek1977projections}.\\

\noindent \textbf{\emph{Dopamine modulation in the network.}} Dopamine neurons from the ventral tegmental area (VTA) innervate many areas including the ventral striatum (nucleus accumbens)~\cite{liddle2000immediate,pan2010inputs}, olfactory bulb, amygdala, hippocampus, orbital and medial prefrontal cortex~\cite{cetin2004dopamine}, and the cingulate cortex. In turn, the VTA receives glutamate and GABA inputs from a wide net of areas (including cortical and limbic areas)~\cite{russo2013brain}.  Excitatory prefrontal projections to the VTA play an important role in regulating the activity of VTA neurons and the extracellular levels of dopamine (DA) within forebrain regions. Previous investigations have demonstrated that PFC terminals synapse on the dendrites of DA and non-DA neurons in the VTA~\cite{carr2000projections}.

In addition to these well established effects of dopamine on reward processing, there are also a few newer finds which complete our knowledge of VTA connectivity, and which we incorporated in our model circuit. The dopamine projection to the thalamus was thought to be non-existent, when a schizophrenia study found high levels of dopamine in the thalamus of patients with normal dopamine levels in striatal regions~\cite{oke1987elevated}. Subsequent work showed that the dopamine projection to the thalamus is not negligible and and incorporated aberrant thalamic dopamine, together with other hypotheses, into a systems model of psychosis~\cite{moghaddam2010dopamine,lisman2010thalamo}.

Finally, there is new evidence that VTA dopamine neurons can also release glutamate or GABA, contributing to wider functional effects~\cite{russo2013brain}. On one hand, dopaminergic neurons were shown to inhibit action potential firing in both direct- and indirect-pathway striatal projection neurons through release of GABA~\cite{tritsch2012dopaminergic}. On the other hand, a recently discovered population of glutamate neurons in the VTA  projects to the nucleus accumbens (NAc), lateral habenula, ventral pallidum (VP), and amygdala~\cite{hnasko2012ventral}.

\subsection{Obsessive compulsive disorder and connectivity in the reward circuit}
\label{hypotheses}

We will build our model upon known aspects of neural circuitry, but, in our future work, we are ultimately planning to compare our model predictions with activation patterns observed via neuroimaging studies at the level of connected regions of interest, in conjunction with different behavioral profiles (anxiety, obsessive thoughts, compulsive behavior). Clearly, the network implicated in reward processing and motion execution is very wide, and a comprehensive model of all its regulatory aspects would be undesirably complex, even when considered in isolation from external interactions, and at a single -- region of interest -- scale. With our model, we choose to focus in particular on a few aspects involved in the dynamics of OCD, and on trying to verify a set of specific hypotheses that have been stated in the current literature in relation to the neurobiological underpinnings of this illness.

A few studies support the broad hypothesis that OCD is associated with functional alterations of brain corticostriatal networks. These findings emphasize that functional connections in the network are modulated by affective and motivational states and further suggest that OCD patients may have modulation abnormalities in this network~\cite{jung2013abnormal}. A variety of studies have additionally related OCD symptoms to abnormal connectivity between cortico-striatal modules and the limbic and dopamine systems. This is not surprising, since the latter are responsible for crucial aspects of emotional responses and reward processing, suggesting their potential involvement in the anxiety control whose malfunction may lead to obsessive thoughts and compulsive movement execution. Below are the existing connectivity hypotheses that we consider in our model:\\

\noindent \textbf{\emph{Hypothesis H1: Enhanced OFC projections to the NAc correlate with OCD symptoms.}} An fMRI study comparing 21 patients with OCD and 21 matched healthy controls~\cite{harrison2009altered} found abnormal and heightened functional connectivity of ventrolimbic corticostriatal regions in patientswithOCD. More precisely, patients with OCD had significantly increased functional connectivity along a ventral corticostriatal axis, implicating the orbitofrontal cortex and surrounding areas. The specific strength of connectivity between the ventral caudate/nucleus accumbens and the anterior orbitofrontal cortex predicted patients' overall symptom severity. \\

\noindent \textbf{\emph{Hypothesis H2: Decreased amygdala connectivity with the NAc correlate with OCD symptoms.}} Other studies have confirmed the increased functional connectivity between the nucleus accumbens and the lateral orbitofrontal cortex during resting-state in OCD patients~\cite{jung2013abnormal}. A tractography study showed that in fact this functional abnormality reflects a structural issue: the fractional anisotropy of fibers between the orbitofrontal cortex and the striatum was higher in OCD patients compared to healthy controls~\cite{nakamae2013altered}. Additionally, patients were shown to have decreased connectivity between the nucleus accumbens and limbic areas such as the amygdala during incentive processing~\cite{jung2013abnormal}.\\ 

\noindent \textbf{\emph{Hypothesis H3: Decreased VTA dopamine modulation of the NAc correlate with OCD symptoms.}} new study by Harrison et al. found an unanticipated effect: patients with OCD showed evidence of reduced functional connectivity of the ventral striatum with the ventral tegmental area~\cite{harrison2009altered}. In healthy controls, the VTA provides the ventral striatum (nucleus accumbens) with dense dopaminergic innervation that are presumed to be critical for cortically driven action selection and long-term plasticity of corticostriatal loops~\cite{surmeier2007d1}. While current evidence from in vivo studies suggests that striatal dopamine activity may be elevated in patients with OCD, findings have been mixed~\cite{denys2003role}. In our model, we will test the implications of this hypothesis.\\

\noindent \textbf{\emph{Hypothesis H4: Decreased inhibition of the amygdala increases anxiety.}} In our previous modeling work, we have studied two types of inhibition to the amygdala, in relation with their potential contributions to keeping amygdala activation levels under control: self-inhibition in the amygdala, and indirect inhibition from cortical areas. Decreased inhibition of both types has been related to emotional disturbances such as anxiety and psychosis. We hypothesize that inhibition of the amygdala also plays an important role in the dynamics of OCD symptoms.

%%%%%%%%%%%%%%%%%%%%%%%%%%%%%%

\section{Model construction and hypotheses}
\label{mother_model}

We showed in the introduction that a very complex network of regions is involved in the abnormal reward and emotional processing. There are a few distinct hypotheses on specific abnormal mechanics leading to the type of increased anxiety and drive to motion execution which feed into obsessive and compulsive behavior. Here, we investigate only a few such factors, hence we tailored our system to optimally address these problems. We included in our network the following coupled areas ($O$=orbitofrontal cortex, $C$=cingulate cortex, $A$= amygdala, $T$=thalamus, $S$=ventral striatum=nucleus accumbens, $\delta$=dopamine=ventral tegmental area), which our focus hypotheses hold responsible for the network dynamic abnormalities. This will help keep the system simple enough to be computationally tractable, while still sufficiently comprehensive to illustrate our specific problems. Some extensions in future work are discussed in Section~\ref{discussion}.

To fix our ideas mathematically, we decided to introduce all non-dopaminergic connections as their a first (linear) Taylor approximation term around zero. The dopaminergic connections were introduced as nonlinear (sigmoidal) terms, since the nonlinearity may capture in this case more of the diffuse effect and timing of dopamine at its target sites. 

\begin{eqnarray}
dO/dt &=& -nO+mA+mT + f_\mu(O,\delta)  \nonumber \\
dC/dt &=& mO-nC+mT + f_\mu(C,\delta) \nonumber \\
dA/dt &=& -aO-aC-n_AA+mT + m\delta + f_\mu(A,\delta) \nonumber \\ 
dT/dt &=& mO+mC+mA-nT+mS +f_\mu(T,\delta) \nonumber \\
dS/dt &=& b_1O+b_2A+mT-nS - m\delta + f_\lambda(S,\delta) \nonumber \\
d\delta/dt &=& = m(O+C+A+T+S)-n\delta
\label{mother_sys}
\end{eqnarray}

The linear coefficients and modulations are as follows. The parameter $m$ represents the strength of excitatory control throughout the network and $n$ is the level of self-damping (taken to be uniform for all areas in the network), so that, in absence of all outside stimuli, activity in an isolated area would die out. While clearly the values of $n$ and $m$ are in reality different in each node and respectively between node-pairs, this is a simplification aimed to help with tractability when performing a parameter sensitivity analysis of the system. A too high-dimensional parameter space would make this task practically impossible. This is a typical modeling negotiation between realism and approachability.

The coefficients $b_1$ and $b_2$ are, respectively, the level of orbitofrontal excitatory control of the striatum and  the level of amygdalar excitatory control on the striatum. To illustrate these effects in our model and verify hypotheses \emph{H1} and \emph{H2}, we allowed them to vary within a range of values around what we considered the baseline ($m=1$). One of our goals will be to perform a sensitivity analysis of the model with respect to these two parameters, and verify whether the systemic dynamic effects of increasing $b_1$ and decreasing $b_2$ are consistent with those described in the empirical literature.

The far reaching, complex dopamine modulation on all the areas of the network was represented by the sigmoidal function 
$$f_{\mu}(X,\delta) =  \frac{1}{\ds e^{-\mu(X-\delta)}+1} - \frac{1}{2}$$

This nonlinearly increasing function reflects a plausible biological modulation, with a slow take-off at low dopamine levels, and a window of high-sensitivity followed by saturation. Since it has been suggested that dopamine modulation can have both excitatory and inhibitory effects, we allow both positive and negative values for the function $f_{\mu}$, determined by the level of activation already existent in the target area. The highest sensitivity (position and slope at the inflection point) is tuned by the parameter $\mu$, so that higher $\mu$ values can be interpreted as a stronger dopamine effect. The baseline level for this parameter was considered to be $\mu=0.1$ in all simulations. In order to test hypothesis \emph{H3}, we allowed the effect of dopamine on the nucleus accumbens (and the corresponding sensitivity parameter $\lambda$) to be varied independently from the dopamine contributions to the other areas.

In order to test Hypothesis \emph{H4}, we allowed to test for different levels of amygdalar inhibition. The coefficient $n_A$, representing the strength of self-inhibition within the amygdala, was allowed to vary around the baseline value $n_A=n$. The coefficient $a$, representing the strength of the cortical inhibition of amygdala (assumed to be identical from the orbitofrontal and the cingulate cortices), was allowed to vary in a range around the baseline $a=2$.

%%%%%%%%%%%%%%%%%%%%%%%%%

\section{Results}
\label{results}

\noindent \textbf{\emph{Exploring hypotheses H1 and H2:}} We have investigated the effects on the system's steady state when changing the values of the parameters $b_1$ and $b_2$, and observed that, as hypothesized, both effects (also in combination) are associated with typical obsessive-compulsive symptoms, as predicted by our model. 

For a normal range of $(b_1,b_2)$ values, the system has an attracting spiral, so that, in absence of outside stimulation, activation in all areas of the network decays to a baseline level, to which we will refer as the ``control baseline.'' The value of the control baseline is intrinsically meaningless, and will be only used to interpret, by comparison, the behavior in other parameter regimes (which may deliver a higher or a lower baseline than the control level, or an oscillation around it).

\begin{figure}[h!]
\begin{center}
\includegraphics[width = 0.6\textwidth]{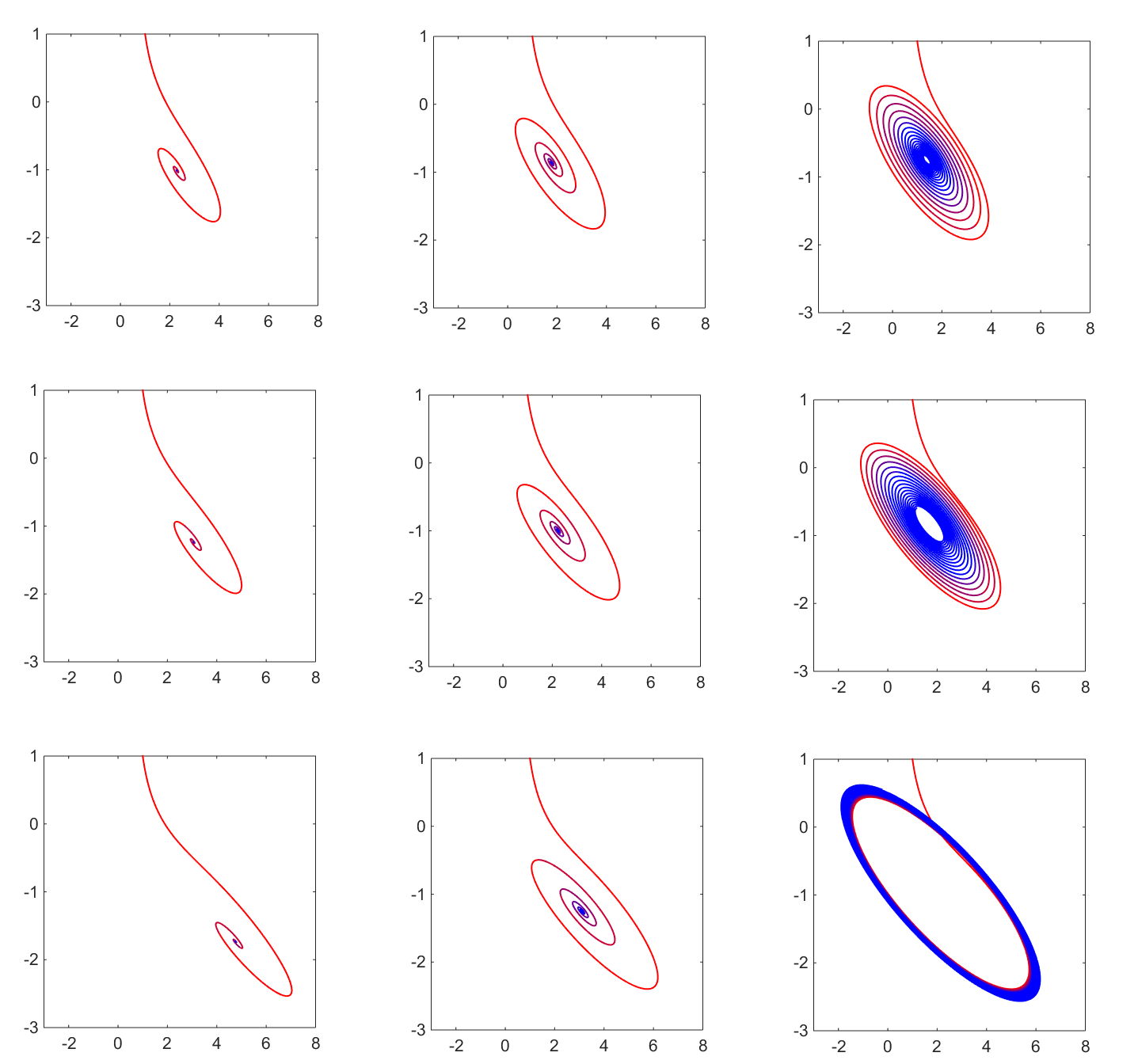}
\end{center}
\caption{\emph {\small {\bf Attracting states of the system for different values of the connectivities $b_1$ and $b_2$.} All panels show the evolution of a typical trajectory in an $(C,A)$ slice (i.e., activation in the amygala on the vertical axis versus activation in the cingulate cortex on the horizontal axis). The simulations were produced for $\lambda=0.1$ (representing lower dopamine modulation of the NAc. Each  row shows a different $b_1$ value: $b_1=0.4$ (top); $b_1=0.8$ (middle); $b_1=1.2$ (bottom). Each column shows a different $b_2$ value: $b_2=0.4$ (left); $b_2=0.8$ (center); $b_2=1.2$ (right). Other parameter values: $m=1$, $n=1.4$, $a=2$, $\mu=0.1$.}}
\label{low_dopa}
\end{figure}

When the parameter $b_2$ is increased, the attracting spiral evolves (through a Hopf bifurcation) towards a stable cycle. This cycle represents a long-term sustained high/low oscillation in the network's activity. In particular, the amygdala and the cingulate cortex oscillate out of phase, so that the network alternates periods of elevated amygdala (high anxiety) with periods of elevated cingulate (high compulsion for motor execution). We will call this an obsessive/compulsive cycle.
  
When the parameter $b_1$ is increased, the system evolves towards a steady state with progressively lower amygdala and higher thalamo-cortical activity. In particular, the cingulate levels increase with $b_1$. We call this regime \emph{compulsive release}, since one may interpret the increased drive for repeated motor execution as anxiety alleviating. 

While the individual effects of $b_1$ and $b_2$ seem relatively clear, the effect on dynamics of these two factors in combination is not trivial to interpret. For small values of $b_1$ (top row in Figure~\ref{low_dopa}), decreasing $b_2$ does not seem to have dramatic consequences, only changing the speed of convergence to the stable equilibrium. For larger $b_1$, however (second and third rows in the same figure), decreasing $b_2$ may transition the system from a range of obsessive/compulsive oscillation to a range of compulsive release. From a different perspective: for larger $b_2$ values (right panel column in Figure~\ref{low_dopa}), increasing $b_1$ pushes the system from a controlled regime into a wide obsessive/compulsive cycle. For smaller values of $b_2$ (left two panel columns in the same figure), increasing $b_1$ pushes the system from a control regime to increased compulsion release.\\

\begin{figure}[h!]
\begin{center}
\includegraphics[width = 0.6\textwidth]{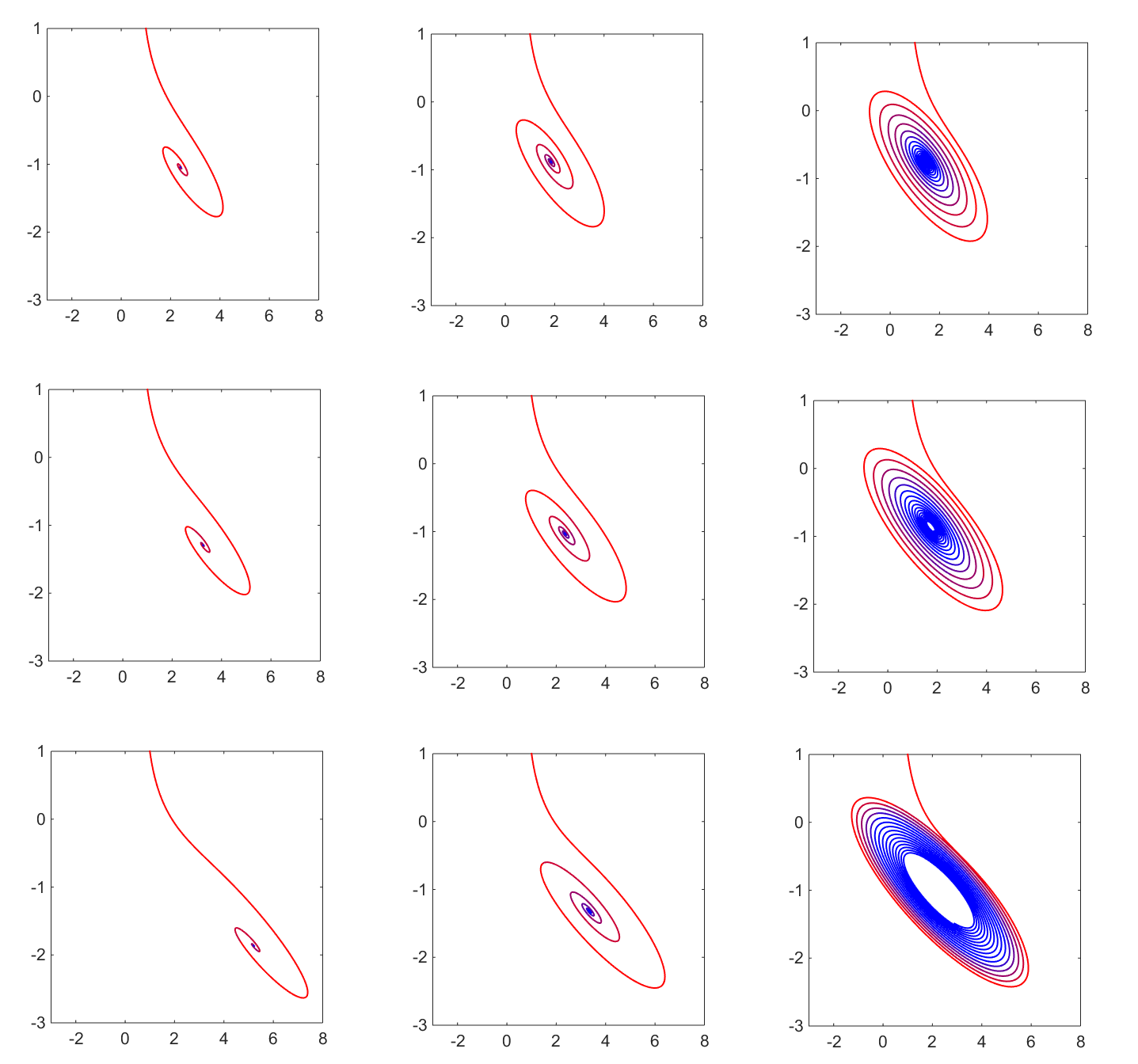}
\end{center}
\caption{\emph {\small {\bf Attracting states of the system for different values of the connectivities $b_1$ and $b_2$.} All panels show the evolution of a typical trajectory in an $(C,A)$ slice (i.e., activation in the amygala on the vertical axis versus activation in the cingulate cortex on the horizontal axis). The simulations were produced for $\lambda=0.2$ (representing higher dopamine modulation of the NAc. Each  row shows a different $b_1$ value: $b_1=0.4$ (top); $b_1=0.8$ (middle); $b_1=1.2$ (bottom). Each column shows a different $b_2$ value: $b_2=0.4$ (left); $b_2=0.8$ (center); $b_2=1.2$ (right). Other parameter values: $m=1$, $n=1.4$, $a=2$, $\mu=0.1$.}}
\label{high_dopa}
\end{figure}

\noindent \textbf{\emph{Exploring hypothesis H3:}} We studied how the behavior described above is altered when changing the dopamine modulation on the nucleus accumbens. We have studied this by increasing the sensitivity $\lambda$ of the sigmoidal modulation used to model the dopamine impact specifically on the variable $S$: for a higher $\lambda$, the sensitivity interval begins at lower activity levels of $S$, allowing dopamine to have a higher impact on this area. This increase dims the effects observed at lower dopamine modulation. In Figure~\ref{high_dopa}, the equivalent of Figure~\ref{low_dopa} for higher dopamine sensitivity of the NAc, one can notice similar -- but milder -- effects and transitions when increasing and respectively decreasing $b_1$ and $b_2$.\\

\begin{figure}[h!]
\begin{center}
\includegraphics[width = 0.8\textwidth]{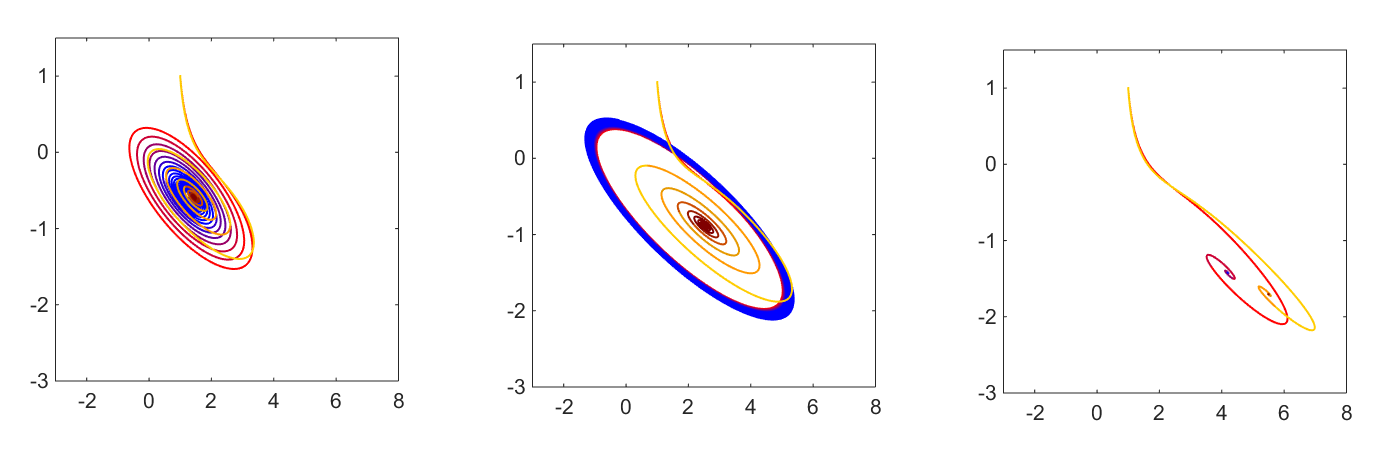}
\end{center}
\caption{\emph {\small {\bf Attracting states of the system for different values of amygdala self-inhibition $n_A$.} In each panel, the evolution of trajectories in the $(A,C)$ slice is shown for lower amygdala self-inhibition $n_A=1.4$ (from red to blue) and for higher $n_A=1.6$. The three different panels represent three different combinations of $(b_1,b_2)$: $b_1=0.4$, $b_2=1.2$ (left); $b_1=1.2$, $b_2=1.2$ (center); $b_1=1.2$, $b_2=0.4$ (right). Other parameter values: $m=1$, $n=1.4$, $a=2$, $\mu=\lambda=0.1$.}}
\label{comparison_n1}
\end{figure}

\noindent \textbf{\emph{Exploring hypothesis H4:}} In previous work, we suggested that amygdalar self-inhibition is an important factor providing stability to the dynamics of the emotion regulatory network~\cite{radulescu2008schizophrenia,radulescu2009multi}. Here, we again observed the effect of increasing the amygdalar self-inhibition $n_A$, for different connectivity profiles $(b_1,b_2)$. We noticed that, for low values of $b_1$ and high values of $b_2$ (corresponding, based on our previous considerations, to a well-controlled system), a slight increase in $n_A$ will only allow trajectories to converge faster, but not affect much the position of the stable equilibrium (Figure~\ref{comparison_n1}a). We observed more substantial effects, however, if the system is already operating in an OCD-prone regime, with a large $b_1$ or/and a low $b_2$.  For high $b_1$ and high $b_2$, increasing $n_A$ brings the system from a range in which the trajectories converge to a stable cycle -- back through a Hopf bifurcation to a regime of relatively fast convergence to a stable equilibirium, at relatively low activation levels of $A$ and $C$ (Figure~\ref{comparison_n1}b). For high $b_1$ and low $b_2$, increasing $n_A$ results in a decrease in the asymptotic amygdala levels, at the expense of increasing cortical activation (less anxiety, but more pronounced compulsion, Figure~\ref{comparison_n1}c).

\begin{figure}[h!]
\begin{center}
\includegraphics[width = 0.8\textwidth]{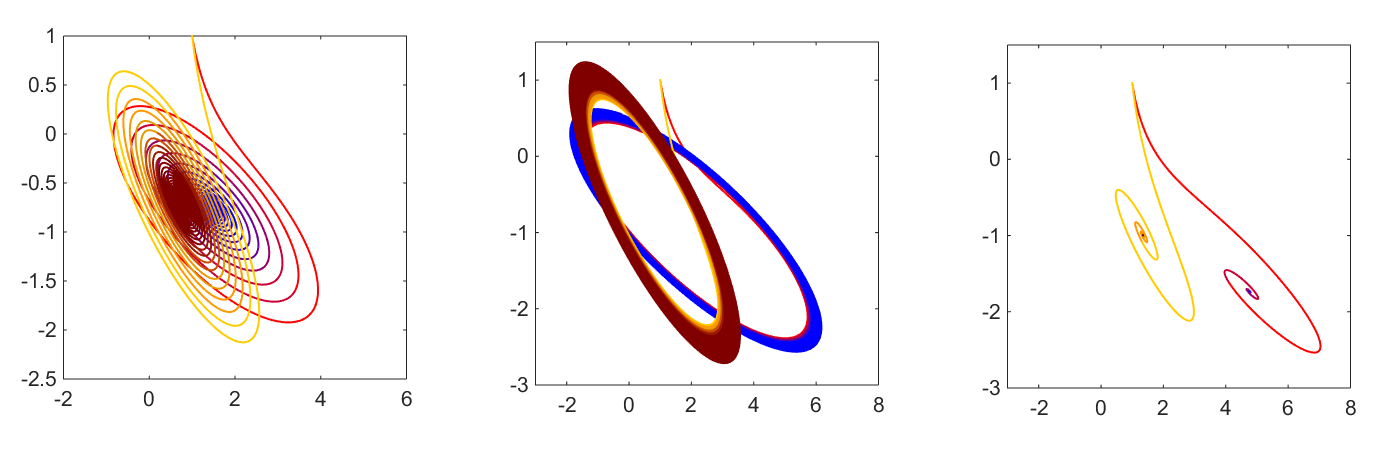}
\end{center}
\caption{\emph {\small {\bf Attracting states of the system for different values of the cortical inhibition of the amygdala $a$.} In each panel, the evolution of trajectories in the $(A,C)$ slice is shown for lower inhibition $a=2$ (from red to blue) and for higher inhibition $a=2.5$. The three different panels represent three different combinations of $(b_1,b_2)$: $b_1=0.4$, $b_2=1.2$ (left); $b_1=1.2$,$b_2=1.2$ (center); $b_1=1.2$, $b_2=0.4$ (right). Other parameter values: $m=1$, $n=n_A=1.4$, $\mu=\lambda=0.1$.}}
\label{comparison_a}
\end{figure}

We also studied the effect of increasing the cortical inhibition of the amygdala, represented in our model by the connectivity strength $a$ between orbitofrontal/cingulate areas and the amygdala. While one would expect a stronger inhibition of an area to result in lower activation levels in that respective area, the actual effect produced by the coupling in the network was rather counterintuitive: is was the levels of cortical activation that were in fact reduced as an effect of increasing amygdalar inhibition. In the case of a well controlled network (small $b_1$ and large $b_2$),  lower cingulate activation (shown in Figure~\ref{comparison_a}a) resulted without any detriment to the amygdala function. However, in the case of a larger $b_1$ and/or a lower $b_2$, arousal levels actually increased by increasing $a$, either as a increase of the amygdala steady state (Figure~\ref{comparison_a}c), or as an oscillation with increased amplitude in the amygdala component (Figure~\ref{comparison_a}b).

%%%%%%%%%%%%%%%%%%%%%%%%%%%%%%

\section{Discussion}
\label{discussion}

In this study, we illustrated how a mathematical model of connectivity in a brain network can be used to study the phenomenology of OCD. We focused primarily on investigating how specific connectivity hypotheses of OCD reflect on behavioral dynamics, as per our modeling setup. We found that different known abnormalities of the reward/executive network affect differently the network dynamics, predicting different types of behavioral regimes associated with OCD. Here, we would like to discuss our analytical findings and put them in the broader perspective of their meaning and relevance to the study of OCD.

Based on its known activation during emotional arousal, we interpret high amygdala as high level of arousal / anxiety associated with the obsessive aspect of the OCD symptomatology. Based on its crucial role in initiating movement execution, we interpret high cingulate cortex activation as the elevated executive drive that  in our case relates to the compulsive aspect of the OCD symptomatology. The different regimes that we identified in our network reflect these behavioral characteristics.

We found one potential asymptotic regime with  a low amygdala and high cingulate steady state. A plausible behavioral interpretation is that, in such as regime, the compulsive movement execution redirects to some extent the obsessive thoughts, and diffuses the anxiety. In more lay terms, the compulsion execution is calming. Second, there is the counterpart regime in which the obsessive thoughts build up into high anxiety (high amygdala) state, with no release coming from movement execution (low cingulate). A third regime is an oscillation between these two ends  (prevalent in OCD~\cite{foa1995dsm}), with obsessive periods followed by compulsive periods, with their duration and severity (duty cycle), depending itself on the specific hardwiring (parameters) of the system.

In particular, we showed that increased orbitofrontal--nucleus accumbens (OFC-NAc) connectivity leads, as hypothesized, to more pronounced OCD featues. The precise type and strength of this relationship additionally depend on the ranges of other parameters. For example, if in conjunction with a high amygdala-striatal (A-NAc) connectivity, increasingly higher values of OFC-NAc will push the system into gradually wider oscillations between obsessive and compulsive periods. In conjunction with a low A-NAc,  increasingly higher values of OFC-NAc will drive the system towards a preferentially compulsive state. These findings are generally in agreement with both hypotheses \emph{H1} and \emph{H2}, although they also suggest that the dependence on these two parameters may be more complex than currently described in empirical studies.

We noticed that a more sensitive striatal dependence on dopamine may regulate the system, diminishing, although not completely eliminating, the OCD signs. This is in agreement with hypothesis \emph{H3}, associating OCD with a weaker dopamine modulation of the ventral striatum.

Finally, we further investigated the role of the amygdala, a key regulator of arousal, in obsessive-compulsive mechanics. A classical and  intuitive view would be that a weaker inhibition of the amygdala (either internal or from external sources, such as cortical areas) would result in higher amygdalar activation, and subsequently in higher anxiety. We found that to be a rather simplistic view, and that, as expected, the effect depends on contributions from other areas, and cannot be judged in absence of more global information on the system. 

For example, when operating in a ``normal'' range of OFC-NAc and A-NAc connectivity, increasing amygdalar self-inhibition has a stabilizing effect (e.g., produces faster convergence of trajectories to a well-controlled equilibirum). This effect is even more dramatic when the system is operating in a range with abnormally high OFC-NAc connectivity, but normal A-NAc connectivity, where a similar increase in amygdala self-inhibition can stop an obsessive-compulsive oscillation and stabilize it to a stable equilibrium within the normal functional range. For high OFC-NAc connectivity in conjunction with low A-NAc connectivity (both described in the literature as factors contributing to OCD symptoms), increasing amygdalar self-inhibition increased ACC asymptotic levels and enhanced quite dramatically the compulsion behavioral aspect.

Similarly, when increasing the cortical inhibition of the amygdala, the effects also differed between the three situations. When under low OFC-NAc and high A-NAc, increasing inhibition only introduces slight changes in the position of the attractor and the evolution path towards it. For high OFC-NAc and high A-NAc, increased inhibition does not break the obsessive compulsive oscillation, but rather changes its geometry, emphasizing larger oscillations along the amygdala axis and lower amplitude along the cingulate axis (i.e., behavior governed more by mood swings than by variations in compulsive behavior). Finally, for high OFC-NAc and low A-NAc, higher inhibition surprisingly reduces cingulate levels, which slightly increasing amygdala levels (that is, the regime sports higher anxiety levels at the expense of reducing compulsions). This is precisely the opposite effect to that of increasing amygdala self-inhibition under otherwise the same circumstances.

In conclusion, our analysis shows that changes in parameters can be responsible for the system operating in different dynamic ranges.  We found all hypotheses investigated to correspond to realizable scenarios that may lead to triggering OCD behavior or contributing to worsening of symptoms. However, it would be a mistake to consider these factors out of context and  in isolation, since each describes a different aspect of the elephant that fills the room, and only in combination could they shed light on the problem as a whole. In addition, connectivity changes are dynamic and fluid, hence our connectivity parameters should also change along with the system's states -- not being adjusted by hand, but rather allowing the system itself to adjust them as it evolves. It is precisely in this direction that mathematical models can provide crucial support. Obtaining tractable biophysical, data-driven models of the reward/executive network may lead to major advances of our knowledge of the causes of OCD.

 Out study in particular has qualities such as computational tractability, generality (in the sense that similar setups can be used towards testing additional hypotheses about function in this network,  as well as in other regulatory loops). We hope that this framework may help other investigators build upon our approach to further address such questions.

However, the model also has clear limitations. One main qualitative restriction imposed on our system is the linearity of excitatory and inhibitory influences between brain areas. This is only a broad first order approximation, and can be refined in future work. Another limitation of our model is its phenomenological aspect -- part of our current work is directed towards estimating rages of empirical values for the system's parameters, allowing more realistic classifications and predictions of possible behaviors. Finally, our model is conceived at a single spatio-temporal scale, that of regions of interest, with the networked brain areas viewed as black boxes (consistent with the perspective offered by imaging studies). This approach does not address neural activity and synaptic patterns, the biophysical level where the mechanics of OCD actually happens. We are currently working on a neural model of the cortico-striato-thalamo-cortical loop, based on population excitatory/inhibitory interactions. A multi-scale neural model obtained by merging these two complexity levels would incorporate the crucial neurobiology, yet allow, via a neuro-vascular filter, comparison with BOLD empirical data from fMRI scans of OCD patients.

\bibliographystyle{plain}
\bibliography{references}

\begin{thebibliography}{10}

\bibitem{american2003apa}
American~Psychiatric Association et~al.
\newblock Apa (2000).
\newblock {\em Diagnostic and statistical manual of mental disorders DSM (4th
  ed., text revision)}, 2003.

\bibitem{besiroglu2011involvement}
L~Besiroglu, M~Sozen, {\"O}~Ozbebit, S~Avcu, Y~Selvi, A~Bora, A~Atli, O~Unal,
  and MD~Bulut.
\newblock The involvement of distinct neural systems in patients with
  obsessive--compulsive disorder with autogenous and reactive obsessions.
\newblock {\em Acta Psychiatrica Scandinavica}, 124(2):141--151, 2011.

\bibitem{bourne2012mechanisms}
Sarah~K Bourne, Christine~A Eckhardt, Sameer~A Sheth, and Emad~N Eskandar.
\newblock Mechanisms of deep brain stimulation for obsessive compulsive
  disorder: effects upon cells and circuits.
\newblock {\em Frontiers in integrative neuroscience}, 6, 2012.

\bibitem{busatto2000voxel}
Geraldo~F Busatto, Denis~R Zamignani, Carlos~A Buchpiguel, Griselda~EJ Garrido,
  Michael~F Glabus, Euclides~T Rocha, Alex~F Maia, Maria~C Rosario-Campos,
  Claudio~Campi Castro, Sergio~S Furuie, et~al.
\newblock A voxel-based investigation of regional cerebral blood flow
  abnormalities in obsessive--compulsive disorder using single photon emission
  computed tomography (spect).
\newblock {\em Psychiatry Research: Neuroimaging}, 99(1):15--27, 2000.

\bibitem{carr2000projections}
David~B Carr and Susan~R Sesack.
\newblock Projections from the rat prefrontal cortex to the ventral tegmental
  area: target specificity in the synaptic associations with mesoaccumbens and
  mesocortical neurons.
\newblock {\em The Journal of neuroscience}, 20(10):3864--3873, 2000.

\bibitem{cavada2000anatomical}
Carmen Cavada, Jaime Tejedor, Roelf~J Cruz-Rizzolo, Fernando
  Reinoso-Su{\'a}rez, et~al.
\newblock The anatomical connections of the macaque monkey orbitofrontal
  cortex. a review.
\newblock {\em Cerebral Cortex}, 10(3):220--242, 2000.

\bibitem{cetin2004dopamine}
Timur Cetin, Florian Freudenberg, Martina F{\"u}chtemeier, and Michael Koch.
\newblock Dopamine in the orbitofrontal cortex regulates operant responding
  under a progressive ratio of reinforcement in rats.
\newblock {\em Neuroscience letters}, 370(2):114--117, 2004.

\bibitem{cho2013cortico}
Youngsun~T Cho, Monique Ernst, and Julie~L Fudge.
\newblock Cortico--amygdala--striatal circuits are organized as hierarchical
  subsystems through the primate amygdala.
\newblock {\em The Journal of Neuroscience}, 33(35):14017--14030, 2013.

\bibitem{denys2003role}
Damiaan Denys, Joseph Zohar, and HG~Westenberg.
\newblock The role of dopamine in obsessive-compulsive disorder: preclinical
  and clinical evidence.
\newblock {\em The Journal of clinical psychiatry}, 65:11--17, 2003.

\bibitem{elliott2000dissociable}
Rebecca Elliott, Raymond~J Dolan, and Chris~D Frith.
\newblock Dissociable functions in the medial and lateral orbitofrontal cortex:
  evidence from human neuroimaging studies.
\newblock {\em Cerebral cortex}, 10(3):308--317, 2000.

\bibitem{etkin2011emotional}
Amit Etkin, Tobias Egner, and Raffael Kalisch.
\newblock Emotional processing in anterior cingulate and medial prefrontal
  cortex.
\newblock {\em Trends in cognitive sciences}, 15(2):85--93, 2011.

\bibitem{evarts1969motor}
EV~Evarts and WT~Thach.
\newblock Motor mechanisms of the cns: cerebrocerebellar interrelations.
\newblock {\em Annual Review of Physiology}, 31(1):451--498, 1969.

\bibitem{figee2011dysfunctional}
Martijn Figee, Matthijs Vink, Femke de~Geus, Nienke Vulink, Dick~J Veltman,
  Herman Westenberg, and Damiaan Denys.
\newblock Dysfunctional reward circuitry in obsessive-compulsive disorder.
\newblock {\em Biological psychiatry}, 69(9):867--874, 2011.

\bibitem{foa1995dsm}
Edna~B Foa and Michael~J Kozak.
\newblock Dsm-iv field trial: Obsessive-compulsive disorder.
\newblock {\em The American journal of psychiatry}, 152(1):90, 1995.

\bibitem{fudge2002amygdaloid}
JL~Fudge, K~Kunishio, P~Walsh, C~Richard, and SN~Haber.
\newblock Amygdaloid projections to ventromedial striatal subterritories in the
  primate.
\newblock {\em Neuroscience}, 110(2):257--275, 2002.

\bibitem{fudge2004amygdaloid}
Julie~L Fudge, Michael~A Breitbart, and Crystal McClain.
\newblock Amygdaloid inputs define a caudal component of the ventral striatum
  in primates.
\newblock {\em Journal of Comparative Neurology}, 476(4):330--347, 2004.

\bibitem{gigg1992glutamatergic}
John Gigg, Aiko~M Tan, and David~M Finch.
\newblock Glutamatergic excitatory responses of anterior cingulate neurons to
  stimulation of the mediodorsal thalamus and their regulation by gaba: An in
  vivo lontophoretic study.
\newblock {\em Cerebral Cortex}, 2(6):477--484, 1992.

\bibitem{graybiel2000toward}
Ann~M Graybiel and Scott~L Rauch.
\newblock Toward a neurobiology of obsessive-compulsive disorder.
\newblock {\em Neuron}, 28(2):343--347, 2000.

\bibitem{haber2008cognitive}
Suzanne~N Haber and Justin~L Brucker.
\newblock Cognitive and limbic circuits that are affected by deep brain
  stimulation.
\newblock {\em Frontiers in bioscience (Landmark edition)}, 14:1823--1834,
  2008.

\bibitem{harrison2009altered}
Ben~J Harrison, Carles Soriano-Mas, Jesus Pujol, Hector Ortiz, Marina
  L{\'o}pez-Sol{\`a}, Rosa Hern{\'a}ndez-Ribas, Joan Deus, Pino Alonso, Murat
  Y{\"u}cel, Christos Pantelis, et~al.
\newblock Altered corticostriatal functional connectivity in
  obsessive-compulsive disorder.
\newblock {\em Archives of general psychiatry}, 66(11):1189--1200, 2009.

\bibitem{hnasko2012ventral}
Thomas~S Hnasko, Gregory~O Hjelmstad, Howard~L Fields, and Robert~H Edwards.
\newblock Ventral tegmental area glutamate neurons: electrophysiological
  properties and projections.
\newblock {\em The Journal of Neuroscience}, 32(43):15076--15085, 2012.

\bibitem{jung2013abnormal}
Wi~Hoon Jung, Do-Hyung Kang, Euitae Kim, Kyung~Soon Shin, Joon~Hwan Jang, and
  Jun~Soo Kwon.
\newblock Abnormal corticostriatal-limbic functional connectivity in
  obsessive--compulsive disorder during reward processing and resting-state.
\newblock {\em Neuroimage: Clinical}, 3:27--38, 2013.

\bibitem{kishi2006topographical}
Toshiro Kishi, Toshiko Tsumori, Shigefumi Yokota, and Yukihiko Yasui.
\newblock Topographical projection from the hippocampal formation to the
  amygdala: a combined anterograde and retrograde tracing study in the rat.
\newblock {\em Journal of Comparative Neurology}, 496(3):349--368, 2006.

\bibitem{krettek1977projections}
JE~Krettek and JL~Price.
\newblock Projections from the amygdaloid complex to the cerebral cortex and
  thalamus in the rat and cat.
\newblock {\em Journal of Comparative Neurology}, 172(4):687--722, 1977.

\bibitem{kringelbach2004functional}
Morten~L Kringelbach and Edmund~T Rolls.
\newblock The functional neuroanatomy of the human orbitofrontal cortex:
  evidence from neuroimaging and neuropsychology.
\newblock {\em Progress in neurobiology}, 72(5):341--372, 2004.

\bibitem{kwon2003similarity}
JS~Kwon, YW~Shin, CW~Kim, YI~Kim, T~Youn, MH~Han, KH~Chang, and JJ~Kim.
\newblock Similarity and disparity of obsessive-compulsive disorder and
  schizophrenia in mr volumetric abnormalities of the hippocampus-amygdala
  complex.
\newblock {\em Journal of Neurology, Neurosurgery \& Psychiatry},
  74(7):962--964, 2003.

\bibitem{leonard1990childhood}
Henrietta~L Leonard, Erica~L Goldberger, Judith~L Rapoport, Deborah~L Cheslow,
  and Susan~E Swedo.
\newblock Childhood rituals: normal development or obsessive-compulsive
  symptoms?
\newblock {\em Journal of the American Academy of Child \& Adolescent
  Psychiatry}, 29(1):17--23, 1990.

\bibitem{liddle2000immediate}
Peter~F Liddle, Carol~J Lane, and Elton~TC Ngan.
\newblock Immediate effects of risperidone on cortico—striato—thalamic
  loops and the hippocampus.
\newblock {\em The British Journal of Psychiatry}, 177(5):402--407, 2000.

\bibitem{lisman2010thalamo}
John~E Lisman, Hyun~Jae Pi, Yuchun Zhang, and Nonna~A Otmakhova.
\newblock A thalamo-hippocampal-ventral tegmental area loop may produce the
  positive feedback that underlies the psychotic break in schizophrenia.
\newblock {\em Biological psychiatry}, 68(1):17--24, 2010.

\bibitem{malenka2009molecular}
Robert~C Malenka, EJ~Nestler, SE~Hyman, A~Sydor, RY~Brown, et~al.
\newblock Molecular neuropharmacology: A foundation for clinical neuroscience,
  2009.

\bibitem{markarian2010multiple}
Yeraz Markarian, Michael~J Larson, Mirela~A Aldea, Scott~A Baldwin, Daniel
  Good, Arjan Berkeljon, Tanya~K Murphy, Eric~A Storch, and Dean McKay.
\newblock Multiple pathways to functional impairment in obsessive--compulsive
  disorder.
\newblock {\em Clinical psychology review}, 30(1):78--88, 2010.

\bibitem{mcguire1994functional}
PK~McGuire, CJ~Bench, CD~Frith, IM~Marks, RS~Frackowiak, and RJ~Dolan.
\newblock Functional anatomy of obsessive-compulsive phenomena.
\newblock {\em The British Journal of Psychiatry}, 164(4):459--468, 1994.

\bibitem{menzies2008integrating}
Lara Menzies, Samuel~R Chamberlain, Angela~R Laird, Sarah~M Thelen, Barbara~J
  Sahakian, and Ed~T Bullmore.
\newblock Integrating evidence from neuroimaging and neuropsychological studies
  of obsessive-compulsive disorder: the orbitofronto-striatal model revisited.
\newblock {\em Neuroscience \& Biobehavioral Reviews}, 32(3):525--549, 2008.

\bibitem{moghaddam2010dopamine}
Bita Moghaddam.
\newblock Dopamine in the thalamus: a hotbed for psychosis?
\newblock {\em Biological psychiatry}, 68(1):3, 2010.

\bibitem{nakamae2013altered}
T~Nakamae, Y~Sakai, Y~Abe, S~Nishida, K~Fukui, K~Yamada, M~Kubota, D~Denys, and
  J~Narumoto.
\newblock Altered fronto-striatal fiber topography and connectivity in
  obsessive-compulsive disorder.
\newblock {\em PloS one}, 9(11):e112075--e112075, 2013.

\bibitem{oke1987elevated}
Arvin~F Oke and Ralph~N Adams.
\newblock Elevated thalamic dopamine: possible link to sensory dysfunctions in
  schizophrenia.
\newblock {\em Schizophrenia Bulletin}, 13(4):589--604, 1987.

\bibitem{pan2010inputs}
Weixing~X Pan, Tianyi Mao, and Joshua~T Dudman.
\newblock Inputs to the dorsal striatum of the mouse reflect the parallel
  circuit architecture of the forebrain.
\newblock {\em Frontiers in neuroanatomy}, 4, 2010.

\bibitem{radulescu2008schizophrenia}
Anca Raˇdulescu.
\newblock Schizophrenia—a parameters’ game?
\newblock {\em Journal of theoretical biology}, 254(1):89--98, 2008.

\bibitem{radulescu2009multi}
Anca Raˇdulescu.
\newblock A multi-etiology model of systemic degeneration in schizophrenia.
\newblock {\em Journal of theoretical biology}, 259(2):269--279, 2009.

\bibitem{rempel2007role}
NANCY~L REMPEL-CLOWER.
\newblock Role of orbitofrontal cortex connections in emotion.
\newblock {\em Annals of the New York Academy of Sciences}, 1121(1):72--86,
  2007.

\bibitem{richter2000amygdala}
Gal Richter-Levin and Irit Akirav.
\newblock Amygdala-hippocampus dynamic interaction in relation to memory.
\newblock {\em Molecular neurobiology}, 22(1-3):11--20, 2000.

\bibitem{rotge2008provocation}
Jean-Yves Rotge, Dominique Guehl, Bixente Dilharreguy, Emmanuel Cuny, Jean
  Tignol, Bernard Bioulac, Michele Allard, Pierre Burbaud, and Bruno
  Aouizerate.
\newblock Provocation of obsessive--compulsive symptoms: a quantitative
  voxel-based meta-analysis of functional neuroimaging studies.
\newblock {\em Journal of psychiatry \& neuroscience: JPN}, 33(5):405, 2008.

\bibitem{rotge2010gray}
Jean-Yves Rotge, Nicolas Langbour, Dominique Guehl, Bernard Bioulac, Nematollah
  Jaafari, Michele Allard, Bruno Aouizerate, and Pierre Burbaud.
\newblock Gray matter alterations in obsessive--compulsive disorder: an
  anatomic likelihood estimation meta-analysis.
\newblock {\em Neuropsychopharmacology}, 35(3):686--691, 2010.

\bibitem{russo2013brain}
Scott~J Russo and Eric~J Nestler.
\newblock The brain reward circuitry in mood disorders.
\newblock {\em Nature Reviews Neuroscience}, 14(9):609--625, 2013.

\bibitem{skoog1999}
Gunnar Skoog and Ingmar Skoog.
\newblock A 40-year follow-up of patients with obsessive-compulsive disorder.
\newblock {\em Archives of General Psychiatry}, 56(2):121--127, 1999.

\bibitem{sotres2004emotional}
Francisco Sotres-Bayon, David~EA Bush, and Joseph~E LeDoux.
\newblock Emotional perseveration: an update on prefrontal-amygdala
  interactions in fear extinction.
\newblock {\em Learning \& Memory}, 11(5):525--535, 2004.

\bibitem{surmeier2007d1}
D~James Surmeier, Jun Ding, Michelle Day, Zhongfeng Wang, and Weixing Shen.
\newblock D1 and d2 dopamine-receptor modulation of striatal glutamatergic
  signaling in striatal medium spiny neurons.
\newblock {\em Trends in neurosciences}, 30(5):228--235, 2007.

\bibitem{swenson2006review}
Rand~S Swenson.
\newblock Review of clinical and functional neuroscience.
\newblock {\em Educational review manual in neurology. Castle Connolly Graduate
  Medical Publishing, New York}, 2006.

\bibitem{tritsch2012dopaminergic}
Nicolas~X Tritsch, Jun~B Ding, and Bernardo~L Sabatini.
\newblock Dopaminergic neurons inhibit striatal output through non-canonical
  release of gaba.
\newblock {\em Nature}, 490(7419):262--266, 2012.

\bibitem{turner1991thalamoamygdaloid}
Blair~H Turner and Miles Herkenham.
\newblock Thalamoamygdaloid projections in the rat: a test of the amygdala's
  role in sensory processing.
\newblock {\em Journal of Comparative Neurology}, 313(2):295--325, 1991.

\bibitem{vogt1993neurobiology}
Brent~Alan Vogt, Michael Gabriel, et~al.
\newblock {\em Neurobiology of cingulate cortex and limbic thalamus}.
\newblock Springer, 1993.

\bibitem{wall2013differential}
Nicholas~R Wall, Mauricio De~La~Parra, Edward~M Callaway, and Anatol~C
  Kreitzer.
\newblock Differential innervation of direct-and indirect-pathway striatal
  projection neurons.
\newblock {\em Neuron}, 79(2):347--360, 2013.

\bibitem{yager2015ins}
LM~Yager, AF~Garcia, AM~Wunsch, and SM~Ferguson.
\newblock The ins and outs of the striatum: Role in drug addiction.
\newblock {\em Neuroscience}, 301:529--541, 2015.

\end{thebibliography}

\end{document}